\documentclass[twocolumn,prb,showpacs]{revtex4-1}
\usepackage{graphicx}

\begin{document}
\title{Ground state properties of antiferromagnetic anisotropic
   S=1 Heisenberg spin chains}
  \author{D.~Peters}
  \affiliation{Institut f\"ur Theoretische Physik,
    RWTH Aachen,
    52056 Aachen, Germany}
  \author{I.~P.~McCulloch}
  \affiliation{Centre for Engineered Quantum Systems, School of
    Mathematics and Physics, The University of Queensland, 
    St Lucia, QLD 4072, Australia}
  \author{W.~Selke}
  \affiliation{Institut f\"ur Theoretische Physik,
    RWTH Aachen, and JARA-HPC, 52056 Aachen, Germany}

\begin{abstract}

Using (infinite) density matrix renormalization group
techniques, ground state properties of antiferromagnetic S=1
Heisenberg spin chains with exchange and single--site
anisotropies in an external field are studied. The phase
diagram is known to display a plenitude of interesting
phases. We elucidate quantum phase transitions between
the supersolid and spin--liquid as well as the spin--liquid 
and the ferromagnetic phases. Analyzing
spin correlation functions in the spin--liquid phase, 
commensurate and (two distinct)
incommensurate regions are identified.
\end{abstract}

\pacs{75.10.Jm, 75.40.Mg, 75.40.Cx}

\maketitle

In recent years, ground state properties of the 
antiferromagnetic Heisenberg spin--1 chain with single--site 
and uniaxial exchange anisotropies 
in an external field have
been investigated rather
extensively \cite{tone,sen,pet1,pet2,ross}. The
model is described by the Hamiltonian 

\begin{eqnarray}
{\cal H} &=& \sum\limits_{i} 
   (J (S_i^x S_{i+1}^x + S_i^y S_{i+1}^y
    + \Delta S_i^z S_{i+1}^z)\nonumber\\
   \label{Ham}
   & &
   {}+ D (S_i^z)^2  - B S_i^z) 
\end{eqnarray}

\noindent
where $i$ denotes the lattice sites, $\Delta$ the
exchange, and $D$ the single--site anisotropy. The
external field $B$ acts along the $z$--direction.

The magnet displays various intriguing phases at zero
temperature (and, thence, corresponding quantum phase
transitions), the
antiferromagnetic (AF), ferromagnetic (F), half--magnetization
plateau (HMP), spin--liquid (SL), supersolid (SS), and
'large--D' phases. Some of these phases, the AF, F, SL, and SS phases,
show up in the corresponding classical
Heisenberg model \cite{pet1,pet2,holt1}, while the HMP and 
large--D phases reflect the discretization of the spin
orientations in the quantum case. The theoretical efforts have
been motivated and inspired, partly, by related  
experiments \cite{zhou,kim}.

Perhaps most attention, in the context of this, Eq. (1), and
similar \cite{troy,ng,flora,siza,shann,penc} models, has been paid to
the supersolid phase \cite{nuss}, being
the analog of the 'mixed' or 'biconical' \cite{koster} phase in the classical
limit \cite{mat,fish2}. Note that a mapping from 
quantum lattice gases to magnetic systems, explaining the term 
'supersolid' for magnets, has been introduced
some decades ago \cite {matsu}. Typically, quantum fluctuations
tend to reduce substantially the range of stability of the
supersolid phase \cite{pet2,troy}, as compared to the classical 
variant.

The spin correlations in the supersolid phase of the
anisotropic Heisenberg spin chain, (1), have
been argued to behave like in a Luttinger liquid, with
algebraic spatial decay \cite{sen}. Magnetization
profiles have revealed the close analogy of the supersolid to the
corresponding classical biconical structures \cite{pet1,pet2}. The 
critical exponent of the spin stiffness, describing the transition to
the bordering AF and HMP phases, has been found \cite{sen,ross} to
be 1/2.

In this contribution, we shall consider interesting
aspects of the model (1) which
have not been studied in detail so far. We shall
deal with the transition between
the supersolid and the spin--liquid phases as well as with 
the SL--F quantum phase transition. In addition, spin
correlations in the SL phase will be analyzed, to
clarify, especially, previous suggestions on distinguishing commensurate 
and incommensurate regions in that phase \cite{tone,pet1,pet2}. 

\begin{figure}\centering
        \includegraphics[width=220pt,angle=0]{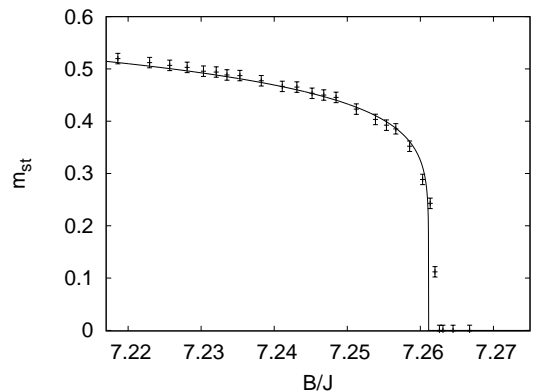}
\caption{\label{fig1} Staggered 
  magnetization $m_{st}$ vs.
  external field $B/J$ at $D/J=\Delta/2= 2.5$ near the supersolid
  to spin--liquid transition. For comparison, a
  power--law fit to the iDMRG data with the
  critical exponent $\beta= 1/8$ is shown, see text.}
\end{figure}

In the present study, mainly infinite density matrix
renormalization group (iDMRG)
techniques \cite {white,scholl,mcc} have been used, with systematic
enlargening on the number of matrices in the matrix product states. In
general, the chosen size of the matrices is an important
parameter determining the reliability of the calculation. Here, the
dimension $M$ of the largest matrices ranges, typically, from
50 to 500. The truncation error varies in between $10^{-6}$
and $10^{-10}$. In a few cases, results are compared to ones we
obtain from DMRG calculations for finite chains, with
open boundary conditions, of length $L$, with
$L \le 128$.

Following previous analyses \cite{tone,sen,pet1,pet2,ross} of
the model (1), we focus
on two cases:  at fixed ratio between the two types of
anisotropies, $D/J= \Delta /2$, and at given, quite large 
exchange anisotropy, $\Delta= 5$, with varying single--site
anisotropy $D/J$. (Parts of) the corresponding 
ground state phase diagrams have been
obtained before \cite{tone,sen,pet1,pet2,ross}, using
DMRG and quantum Monte Carlo techniques. In both cases, the
phase diagrams include the AF, F, HMP, SL, and SS phases. The
supersolid phase results from competing uniaxial, along the $z$--axis,
exchange, $\Delta > 1$, and planar single--site, $D > 0$,
anisotropies.

\begin{figure}\centering
        \includegraphics[width=220pt,angle=0]{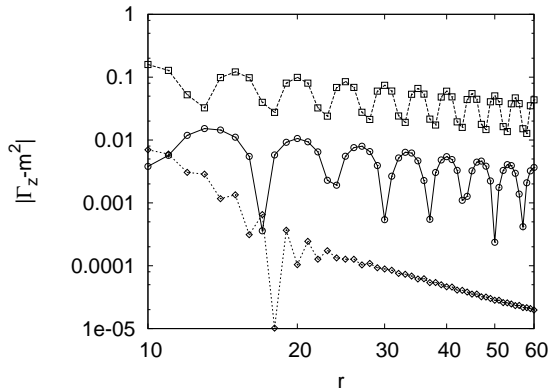}
\caption{\label{fig2} Longitudinal spin correlation function 
  $| \Gamma_z -m^2|$, vs separation distance $r$, at $\Delta=5$, 
  with (a) $D/J= 2.5$, $m= 23/40$ (circles),
  (IC)$_1$ type, (b) $D/J= -1.5$, $m= 1/5$ (squares),
  (IC)$_1$ type, and (c) $D/J= 1$, $m=4/10$ (diamonds), C
  type. Interpolating lines are guides to the eye.}
\end{figure}

The supersolid phase may be bordered by massive, AF or HMP, or by
critical, SL, phases \cite{sen,pet1,pet2,ross}. The transitions
to the massive phases have been
investigated in detail before \cite{sen,ross}. Here we discuss the
transition from the supersolid to the spin--liquid
phase, the SS--SL transition. As illustrated
in Fig. 1 for
$\Delta = 2D/J= 5$, the transition seems to belong to the two--dimensional 
classical Ising universality class. The critical exponent
$\beta$, describing the vanishing of the staggered
magnetization $m_{st}$ at the transition is, indeed, consistent with
the famous Onsager value $\beta =1/8$. Actually, we did
a $\chi^2$--fit of our iDMRG data in the
range $7.217 < B/J < 7.262$ to 
the form $m_{st} = a (B_c/J -B/J)^{1/8}$. We
obtain $B_c \approx 7.261$, as shown in Fig. 1. Note the
(small) deviation extremely close to $B_c$, which one may attribute
to discretization error in determining the magnetic
field or to insufficient size of the matrix dimension $M$, $M \leq 500$, 
in this regime near the transition.  

In addition, we also identify
an Ising--like sector near the SS--SL transition, with
exponentially decaying longitudinal spin correlations up to rather
large distances, as will be discussed below. Note that our suggestion
on the universality class of the SS--SL transition
is in line with a corresponding finding on a related
two--dimensional quantum anisotropic Heisenberg antiferromagnet, where the
supersolid to spin--liquid transition has been
concluded to be in the universality
class of the three--dimensional Ising case \cite{flora}. It is
worth mentioning that both suggestions, for quantum magnets in
dimensions $d$= 1 and 2 at zero temperature, agree
with the well--known dimensional argument \cite{suzu} relating
critical exponents in $d$--dimensional quantum systems to those in
corresponding $(d+1)$--dimensional classical systems. Indeed, here 
the classical transitions between the biconical and
spin--flop phases are of Ising type \cite{koster,holt1}.  

\begin{figure}[h]\centering
        \includegraphics[width=220pt,angle=0]{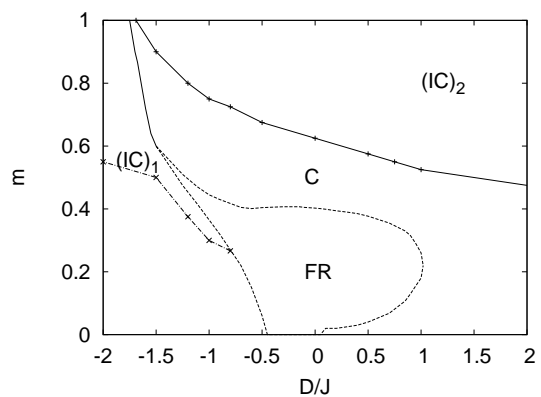}
\caption{\label{fig3}Solid boundary lines separate
  C and IC regions in 
  the $(D/J,m)$ plane at $\Delta= 5$. The dashed line sketches, as
  a guide to the eye, the border of the forbidden
  region (FR), see Ref. 1. The dashed--dotted line in the (IC)$_1$
  phase divides ferroquadrupolar, at high $m$, from
  spin--density--wave ordering.}
\end{figure}
 
Before turning to the discussion of the SL--F quantum
phase transition, let us first consider characteristics of
the SL phase. As has been noted
before \cite{tone,pet1,pet2}, the Hamiltonian (1) may describe
both commensurate (C) and incommensurate (IC) spin--liquid structures, as
has been inferred from the behavior of energy gaps \cite{tone} and
magnetization profiles \cite{pet1,pet2}.

In this study, we shall present direct evidence for both types of
structures in the spin--liquid phase by analyzing, especially, longitudinal
$\Gamma _z(r)= \langle S_i^zS_{i+r}^z \rangle$ spin correlation
functions. Asymptotically, $r \longrightarrow \infty$, $\Gamma_z$ 
acquires the value $m^2$, where $m$ is the total magnetization
per site. For sufficiently large distances, $r$, the dominant
algebraically decaying term of the correlations
may be expected \cite{tsve} to be of
commensurate (C) form, $\propto 1/r^2$, or of
incommensurate (IC) form, $\propto (1/r^{\eta}) \cos(qr)$, with
$\eta < 2$, as usual for Luttinger liquids. Such a behavior is confirmed by
our iDMRG calculations. In the IC case, we find the wavenumber $q$ to
be related to the total magnetization
per site, $m$, in two distinct ways: We obtain either
$q_1 = \pi (1-m)$ ((IC)$_1$), or $q_2 = 2 \pi (1-m)$ ((IC)$_2$), setting
the lattice constant equal to one.

Examples of longitudinal correlation functions of type C, (IC)$_1$,
and (IC)$_2$ are depicted in
Fig. 2, for selected values of $D/J$ and $m$, fixing
the exchange anisotropy, $\Delta =5$. Note that in the example
for the C case, $\Gamma_z$
shows roughly an exponential decay with oscillations at small
distances $r$, approaching the monotonic algebraic 
decay, $\propto 1/r^2$, only at larger separations. 

Varying systematically
the single--site anisotropy $D$ and the
magnetization $m$, at $\Delta=5$ (compare to Ref. 1), one
may then identify three different regions, C, (IC)$_1$, and
(IC)$_2$, in the ($D/J,m$) plane, as shown in Fig.3. At
sufficiently small negative values of $D$, one observes the 
(IC)$_1$ region. This region may be subdivided into two parts: At
larger magnetizations, we observe ferroquadrupolar
ordering \cite{lauch,manna,pet3},
where the algebraic decay of the four--point transverse correlation function 
$\langle (S_i^+)^2(S_{i+r}^-)^2\rangle$ is slower than that of
$\Gamma_z(r)$, due to a smaller exponent $\eta$. At lower magnetizations, one
encounters a spin--density--wave ordering, with the longitudinal
spin correlations being dominant. The (IC)$_2$ region
occurs at larger values of $D/J$. In between the two IC regions, the
commensurate region intervenes. There, the exponent
$\eta$ characterizing the algebraic decay of $\Gamma_z$
with spatially modulated behavior, is larger
than 2. Asymptotically, for large distances
$r$, the dominant algebraic term is then proportional
to $1/r^2$, decaying monotonically. Indeed, the changeover between the C and 
IC regions may be conveniently monitored by determining the
exponent $\eta$ from fits of the iDMRG data for the longitudinal spin 
correlations \cite{pet3}. A few examples, at several fixed values
of $D/J$ and changing $m$, are displayed in Fig. 4, compare to Fig. 3. 

The 'forbidden region' (FR) in the $(D/J, m)$ plane, which has
been sketched in Fig. 3, gives rise to first order transitions.
  
Our calculations on the spin correlations in the 
$(D/J,m)$ plane confirm and refine substantially previous
findings \cite{tone}, where the (IC)$_2$ region seems to
have been overlooked. Similarly, there has been no mentioning
of the two distinct parts of the (IC)$_1$ phase and
of the supersolid phase which shows up at fairly low magnetizations 
and sufficiently large values of $D/J$, as we have
discussed before \cite{pet2}. We omitted the supersolid phase
in Fig. 3, for reasons of simplicity.
 
Close to the supersolid to spin--liquid transition, the
longitudinal spin correlations 
are governed, up to quite large distances, by an
exponential decay with $\pi$--oscillations, signalling, presumably,
the above mentioned
Ising sector. Further details are presented elsewhere \cite{pet3}.

\begin{figure}
  \begin{center}
        \includegraphics[width=220pt,angle=0]{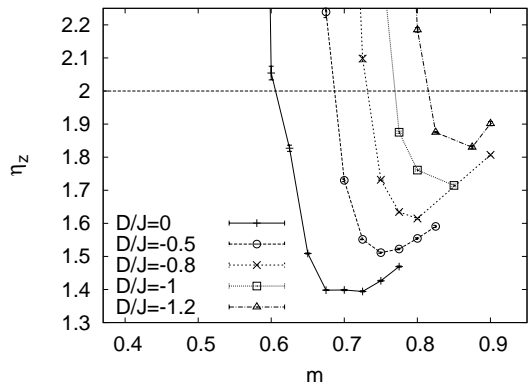}
  \end{center}
  \caption{\label{fig4} Exponent $\eta _z$ for the oscillatory part
   of the algebraically decaying longitudinal spin
   correlations $\Gamma_z$ vs. magnetization $m$ 
   at $\Delta= 5$ and various values of $D/J$.}
\end{figure}

We now turn to the discusssion of the SL--F transition. The
phase transition  may be characterized by the
behavior of the total magnetization per site $m$, as
illustrated in Fig. 5 for $D= 0$, with the SL phase being of
(IC)$_2$ type. Obviously, the
(i--)DMRG data may well be fitted to the form

\begin{equation}
1-m \propto (B_c/J-B/J)^{1/2}
\end{equation}

\noindent
where $B_c$ is the critical field of the SL--F transition. Indeed, we
did a $\chi^2$--fit of the (i)DMRG data in the range between
$11.7 \lesssim B/J \lesssim 12.0$ to Eq. (2), determining
the proportionality factor and the critical field $B_c \approx
12.0$. Further away from the transition, deviations from
the simple power law may be observed, see Fig. 5. Note
that in the (IC)$_2$ region of the SL phase, one has 
$1-m \propto q$. Thence, Eq. (2) corresponds to the 
well known Pokrovsky--Talapov \cite{pokro} square--root
power law for the wavenumber $q$, describing the C--IC transition
in two--dimensional classical systems with uniaxial
spatial anisotropy. Indeed, it seems 
tempting and reasonable \cite{suzu,schulz} to associate the SL--F transition
for the quantum spin chain with that
universality class.

\begin{figure}
  \begin{center}
        \includegraphics[width=220pt,angle=0]{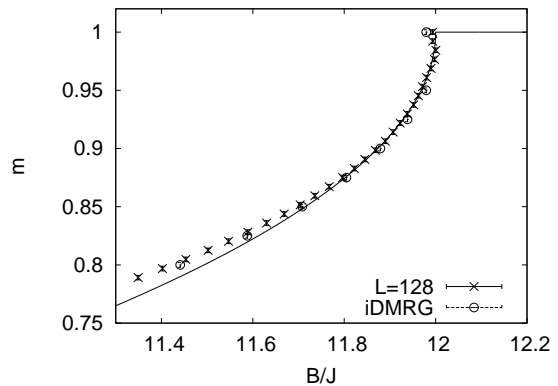}
  \end{center}
  \caption{\label{fig5} Magnetization $m$ versus field $B$ close
   to the SL--F transition at $\Delta= 5$ and $D=0$. Data from iDMRG
   and finite--size DMRG, for $L$= 128 sites, calculations are shown, together
   with a square--root power law fit (solid line), see (2).}
\end{figure}

In summary, studying ground state properties of the S=1 anisotropic
Heisenberg antiferromagnetic chain, using mainly the iDMRG approach, we
present evidence for the spin--liquid to supersolid transition
being in the two--dimensional Ising and for the SL-F transition being in the
Pokrovsky--Talapov universality classes. The spin--liquid phase
is found to consist of commensurate and two distinct incommensurate
regions.

\acknowledgments
We thank Fabian Essler, Frank G\"{o}hmann, Mukul Laad, Salvatore
Manmana, Frederic Mila, Dirk Schuricht, and Stefan Wessel for
very useful discussions.

\end{document}